\documentclass[a4paper,fleqn,usenatbib]{mnras}

\usepackage[T1]{fontenc}
\usepackage{ae,aecompl}

\usepackage{graphicx}	
\usepackage{amsmath}	
\usepackage{amssymb}	
\usepackage{tikz}

\title[Automated Detection of Bars in Galaxies]{Detection of Bars in Galaxies using a Deep Convolutional Neural Network}

\author[Abraham et. al. 2017]{
Sheelu Abraham,$^{1}$\thanks{E-mail: sheelu@iucaa.in}
A. K. Aniyan,$^{2,3}$\thanks{E-mail: arun@ska.ac.za} Ajit K. Kembhavi,$^{1}$\thanks{akk@iucaa.in} N. S. Philip,$^{4}$\thanks{nspp@associates.iucaa.in} \newauthor
Kaustubh Vaghmare $^{1}$\thanks{kaustubh@iucaa.in}
\\
$^{1}$Inter-University Centre for Astronomy and Astrophysics, IUCAA Pune, India\\
$^{2}$Department of Physics and Electronics, Rhodes University, Grahamstown, South Africa\\
$^{3}$SKA South Africa, Pinelands,Cape Town 7405, South Africa\\
$^{4}$Department of Physics, St. Thomas College, Kozhencherry, Kerala, India}
\begin{document}
\label{firstpage}
\pagerange{\pageref{firstpage}--\pageref{lastpage}}
\maketitle

\begin{abstract}
We present an automated method for the detection of bar structure in optical images of galaxies using a deep convolutional neural network which is easy to use and provides good accuracy. In our study we use a sample of 9346 galaxies in the redshift range 0.009-0.2 from the Sloan Digital Sky Survey, which has 3864 barred galaxies, the rest being unbarred. We reach a top precision of ~94 per cent in identifying bars in galaxies using the trained network. This accuracy matches the accuracy reached by human experts on the same data without additional information about the images. Since Deep Convolutional Neural Networks can be scaled to handle large volumes of data, the method is expected to have great relevance in an era where astronomy data is rapidly increasing in terms of volume, variety, volatility and velocity along with other V's that characterize big data. With the trained model we have constructed a catalogue of barred galaxies from SDSS and made it available online. 
\end{abstract}




\section{Introduction}
With the advent of large surveys and facilities for transferring and archiving huge volumes of data, astronomy research has already entered the big data paradigm. In this context, Deep Neural Networks \citep{hinton2006, bengio2009learning, lecun2015deep} which can be scaled to handle large volumes of data have great relevance.  We demonstrate an application of the method for the detection of bar like structures seen in many disc galaxies. 

It has been observed that a significant fraction of disk galaxies in the near Universe have bars \citep{1999ASPC..187...72K}.  Also, hydrodynamical simulations indicate that bars have a clear impact on driving the evolution of the host galaxy by transporting material between the disc and the bulge and thereby redistributing the angular momentum of baryonic and dark matter components of disc galaxies \citep{1985MNRAS.213..451W, 2003MNRAS.341.1179A}. As a consequence, bars play a significant role in the secular evolution of the host disc galaxy \citep{2004ARA&A..42..603K, 2014RvMP...86....1S}. A wider understanding of the prevalence of bars as a function of galaxy type and environment would therefore be of great significance towards our understanding of the evolution of galaxy morphology from early epochs to the present. 

There are multiple methods to detect bars in galaxies. Visual classification is one of the widely accepted methods, but it suffers significantly from subjective biases of the observer and becomes very time consuming with large samples. Another method which works well in the local Universe is to fit ellipses to galaxy isophotes and to detect bars from the peculiarities in their position angles and ellipticity profiles \citep{2003ApJ...592L..13S, 2009A&A...495..491A, 2016A&A...595A..67C}. This method again cannot be scaled to large samples due to the inherent dependence of the method on human interventions. Multi-component image decomposition also helps to identify the presence of a bar structure, but it requires a substantial amount of computation time and human involvement to produce reliable results. 

As an alternative to human expertise, machine learning techniques can be used to address classification and regression problems. The use of machine learning in astronomy is not new. Morphology classification  of galaxies using Artificial Neural Networks (ANN) has been explored by several authors\citep{1992MNRAS.259P...8S, 2010MNRAS.406..342B, 1995MNRAS.275..567N, 2008MNRAS.383..907B}. However, most of the algorithms used until recently depended on a handful of features extracted from the data by human experts for performing the classification. While experts who have gone through years of observation could specify the most reliable features for optimal classification, such specificity is always at the cost of completeness and accuracy on less obvious and finer differences within the same classes.

Moreover, due to the inherent limitation of many of these tools to scale up to handle large volumes of data and feature vectors, the huge volumes of data becoming available from current and forthcoming large-scale surveys such as Sloan Digital Sky Survey  \citep[SDSS;][]{2000AJ....120.1579Y}, SkyMapper \citep{2007PASA...24....1K}, Very Large Telescope Survey Telescope \citep[VST;][]{1998Msngr..93...30A}, Large Synoptic Survey Telescope \citep[LSST;][]{2017arXiv170804058L} will go outdated before they are processed. In addition to that, due to the large variance that is to be expected within the same class, conventional "standard" or "good features" need not be the best representative features for all the samples.

A more practical and desirable approach thus would be to use all available information that can be extracted from the image for doing the classification. This is exactly what Deep Learning methods do, where the input to the classifier is the actual observation itself. Using a convolutional neural network \citep{lecun1995convolutional}, all possible combinations of features are filtered out and used by the classifier to learn the intrinsic differences within the images for attaining highly reliable prediction accuracy on unseen, but similar data. For this reason, in the past few years, Deep Learning Neural Networks have emerged as a highly dependable technique to drive large scale learning problems in astronomy and other branches of science. The availability of extensive data and fast computing systems have further accelerated the development of faster and newer algorithms for this purpose.

In a parallel development, over the last several years, citizen science projects have helped astronomers to overcome the shortage in experts for intensive data analysis. For example, the Galaxy Zoo 2 \citep{2013MNRAS.435.2835W} project with public participation could do sixteen million morphological classifications on 304122 galaxies from  SDSS \citep{2000AJ....120.1579Y}. Results with substantial precision have been obtained from such projects and more importantly, the projects have helped in the discovery of objects with unusual features which would have been difficult to find through conventional methods. But, as stated before, the volume of astronomical data obtained in present and upcoming surveys is growing at such a fast rate that it is impossible to cope with the data through visual classifications. As a result, image processing algorithms which incorporate machine learning are an alternative method for addressing classification problems \citep{2015MNRAS.450.1441D, 2017MNRAS.464.4463K}. In the present study, we augmented classification done by citizen science projects to existing catalogs produced by domain experts to generate training and testing samples for our classifier. The strength of such methods are explored in \citep{2015MNRAS.450.1441D}. The classifier is used to construct a catalogue of 25781 barred galaxies from SDSS DR13. The catalogue has been made it available through a web interface.

The paper is organised as follows: we explain the sample data used in our study in Section~\ref{sec:data} and introduce the Convolutional Neural Network in Section~\ref{sec:CNN}. We describe the network model and the data 
augmentation methods in Sections~\ref{sec:Network} and \ref{sec:Augm} respectively and the training and testing procedures in Section~\ref{sec:Train}. We describe the network analysis and occlusion test in Section~\ref{sec:occ}, our results in Section~\ref{sec:res}, the catalogue and web interface in Section~\ref{sec:cat} and offer concluding remarks in Section~\ref{sec:concl}.

\section{Galaxy Images}\label{sec:data}
In this work, we have used galaxy images from SDSS Data Release 13 (DR13; \citealt{2016arXiv160802013S}) to train the network. SDSS DR13 contains all the observations made upto July, 2015 and is the first data release of SDSS-IV. It provides photometric data in five  bands, \textit{u, g, r, i,} and  \textit{z} covering an area of 14,555 square degrees \citep{1996AJ....111.1748F, 2010AJ....139.1628D}. SDSS Casjobs allows one to access catalogued data measurements from images and spectra such as magnitude, spectral indices, classification, redshifts etc. from different data releases. 

We have used a supervised learning algorithm for barred galaxy classification, so it is important to identify  appropriate training data for the network which includes both barred and unbarred galaxies. Using SDSS Casjobs, we first selected galaxies from DR13 which satisfied the following criteria: 
\begin{itemize}
\item The extinction corrected $r$-band Petrosian magnitude is between 14 and 17.77 because the latter is the faint limit for completeness of the SDSS spectroscopic survey \citep{2000AJ....120.1579Y, 2002AJ....123..485S}.
\item The photometric pipeline identified the object as a galaxy, i.e., type = 3.
\item The spectroscopic class is \textit{GALAXY}.
\item The objects have a clean photometry flag (clean = 1) with no warning flag in spectroscopic measurements (zWarning = 0).
\item The galaxies have half-light radius, measured by the de Vaucouleurs and exponential light profiles, in $r$-band between 5 and 30 arc seconds. 
\end{itemize}

For the barred sample, we choose galaxies identified as barred by human experts, for which we use three catalogues: \citet{2010ApJS..186..427N, 2011MNRAS.415.3627H} and Galaxy Zoo \citep[GZ2;][]{2013MNRAS.435.2835W}. We have cross-matched our sample galaxies with the above mentioned catalogues to see which of our sample galaxies have been identified by them as being barred. See Table~\ref{tab:datasamp} for the barred samples used in this study. We find that from \citet{2010ApJS..186..427N}, 1775 of our galaxies are barred. We visually inspected this set and discarded those galaxies where the bar appeared to be very diffuse or faint, or was very short, or where obvious artefacts were present. This left us with a sample of 776 galaxies. \citet{2011MNRAS.415.3627H} measured the bar length and studied the properties of $\sim$3100 galaxies in the local Universe from Galaxy Zoo data release 2. Application of our selection criteria provides 2308 barred galaxies from their sample of which we selected 1465 via visual inspection in which we removed those galaxies which have a high concentration of artefacts. Lastly, we have selected galaxies which have been identified as barred in the main spectroscopic galaxy sample from the GZ2. There are 6111 galaxies which satisfy our criteria. Excluding galaxies which were already selected from the other two catalogues, we selected only 1623 barred galaxies from GZ2. Thus our final sample consists of 3864 barred galaxies. For the unbarred sample, we selected 5482 galaxies which satisfy our selection criteria and have not been classified as barred by \citeauthor{2010ApJS..186..427N} and GZ2. We visually checked the unbarred galaxies before including them in the final sample.

\begin{table}
\centering
\begin{tabular}{|c|c|c|} \hline 
 & \multicolumn{2}{c|}{Barred Galaxies} \\ \hline
Data & Matched with & Images included \\
& selection criteria & in training set \\ \hline
\citet{2010ApJS..186..427N} & 1775 & 776 \\ \hline
\citet{2011MNRAS.415.3627H} & 2308 & 1465 \\ \hline
\citet{2013MNRAS.435.2835W} & 6111 & 1623  \\ \hline
Total & 10194 & 3864 \\ \hline
\end{tabular}
\caption{Training samples selected for this work.} \label{tab:datasamp}
\end{table}

\begin{figure}
\begin{center}
\includegraphics[scale=0.9]{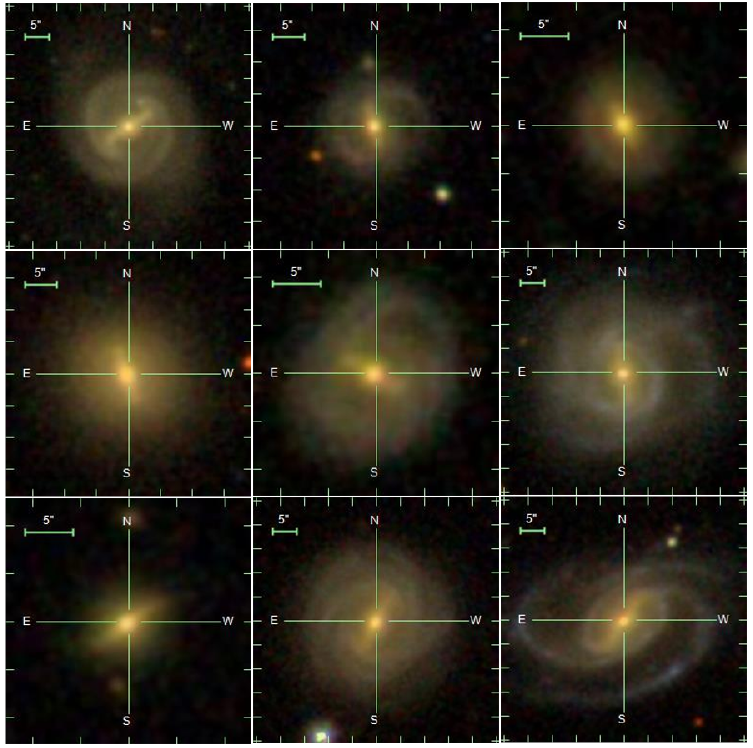}
\caption{Some of the barred galaxy images from SDSS DR13 which are used for the training.}
\label{fig:samp}
\end{center}
\end{figure}

For training our network, we have used JPEG images of galaxies provided by the SDSS $ImgCutout$ service. SDSS applies a $MATLAB$ code to convert the corrected FITS frames of $g, r, i$-band data into 3-colour Joint Photographic Experts Group (JPEG) images. The conversion is based on \citet{2004PASP..116..133L}, further details about which can be obtained from the SDSS website\footnote{http://www.sdss.org/dr13/imaging/jpg-images-on-skyserver/}. The images are of size $300 \times 300$ pixels. We have not applied any other pre-processing on these images. Figure~\ref{fig:samp} shows some example barred galaxies in our training sample.

\section{Convolutional Neural Networks}\label{sec:CNN}

\begin{figure}
 \includegraphics[scale=.25]{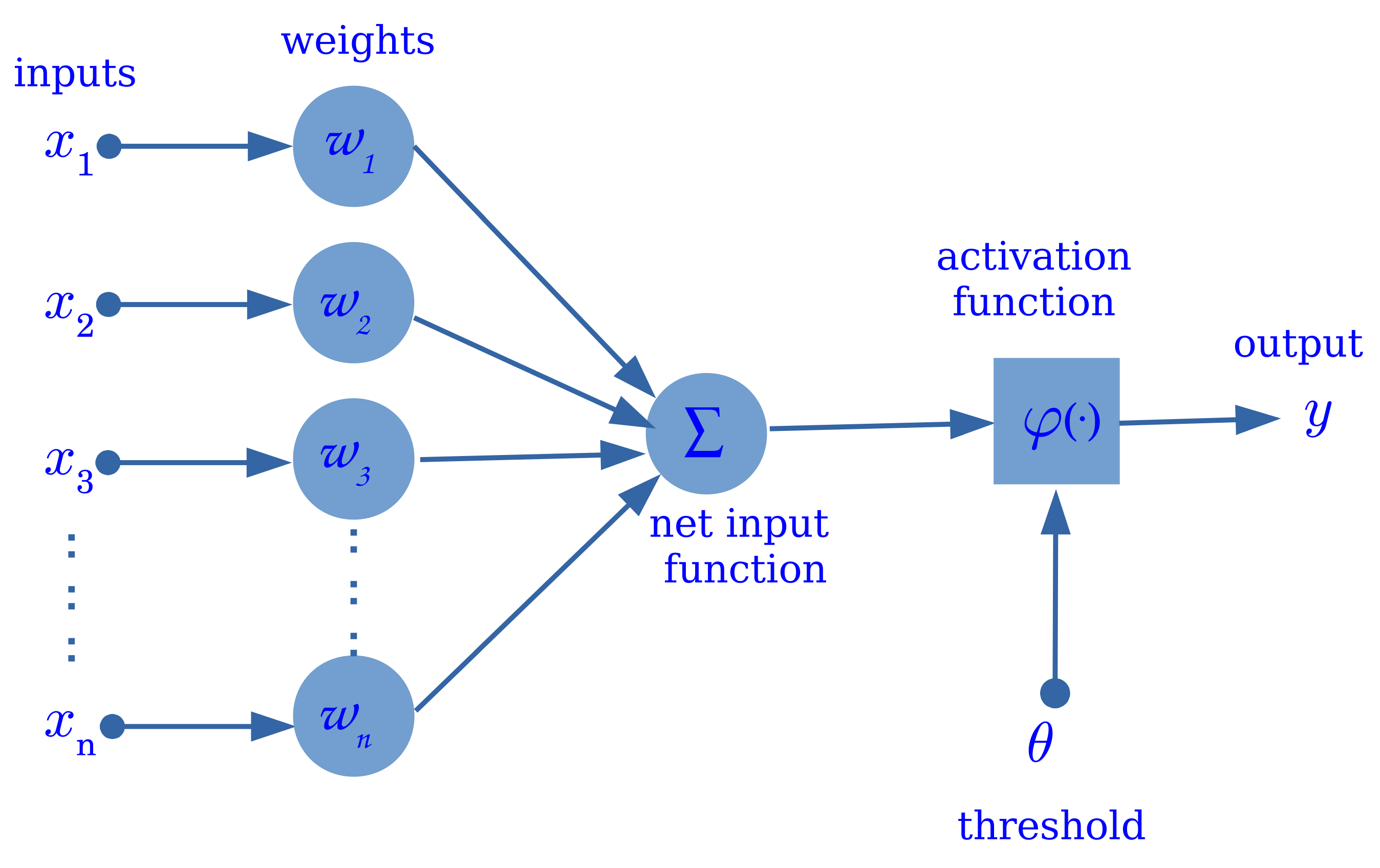}
 \caption{A single neuron with multiple inputs. The figure shows different inputs $[x_1,x_2,..,x_n]$ being weighted by corresponding weights $[w_1,w_2,..,w_n]$ and summed and passed through an activation function to generate the final output $y$.}
 \label{fig:neuron}
 \end{figure}

Artificial neural networks (ANN) were one of the first techniques used in machine learning \citep{duda2012pattern}. ANNs try to mimic biological neural networks which consist of interconnected special cells called neurons. An artificial neuron takes in multiple weighted inputs and generates a summed output, similar to a biological neuron with dendrites which receives input signals, with the resultant output coming out of an axon.  
A simple neuron can be mathematically represented as \citep{duda2012pattern}

\begin{eqnarray} \label{neuron}
y = \sum_{j=1}^{n} w_{j}x_{j} + w_{0},
\end{eqnarray}
where $\mathit{y}$ is the output the neuron with inputs $\mathit{x_{j}}$, weights  $\mathit{w_{j}}$ and bias value $\mathit{w_{\rm 0}}$. A network of such neurons can be interconnected and used for classification and regression applications. The objective function is to find the optimal value of $\mathit{w_{j}}$ which give the desired output $\mathit{y}$. In Figure~\ref{fig:neuron}, we have shown this schematically.   
In a typical feed forward neural network, the inputs $\mathit{x_{j}}$ are feature vectors each of which has an associated  weight $w_{j}$. Such a network can be used as a linear classifier and is often referred as a perceptron \citep{duda2012pattern}. Practical applications require ANNs with multiple layers of neurons called hidden layers with non-linear output functions, in between the input and output layers, to perform classification or regression. During training the sum of the input vectors multiplied with their weights and the bias value are propagated from the input layer to the output layer. This is referred to as forward pass or forward propagation.  At the output layer, the error is calculated between the network output (y) and expected result ($\rm \hat{y}$) as  
\begin{eqnarray} \label{error}
\rm E = \frac{1}{2} || \hat{y} - y||^{2},
\end{eqnarray}
which is send back to the input layers to adjust the weights so as to decrease the error.  This is called a backward pass or backpropagation \citep{rumelhart1986learning,hecht1989theory}.  

Yan LeCun \citep{lecun1995convolutional} first introduced Convolutional Neural Networks (CNN) which are designed to handle data in its one/two/three-dimensional raw form. It was a major breakthrough in the area of computer vision. CNNs made a radical shift in machine learning to a technique in which the machine learning algorithm learned by extracting features automatically \citep{humphrey2012moving}. It was initially applied to images and later found broad applications to many fields including text and speech signal processing   \citep{hinton2012deep,kalchbrenner2014convolutional}. In CNN, features are automatically learned by convolving the input data with filters whose weights are adjusted through backpropagation. For an input image $\mathit{I}$ this can be mathematically represented as \citep{stutz2014understanding}
\begin{equation}
\label{eq:convolution}
\left(I \ast K\right)_{r,s} := \sum _{u = -h_1} ^{h_1} \sum _{v = -h_2}^{h_2} K_{u,v} I_{r+u,s+v},
\end{equation}
where $\mathit{h_{1}}$ and $\mathit{h_{2}}$ indicate the size of the filter $\mathit{K}$ that is learned. When applying CNNs to different classification tasks multiple $\mathit{K}$ filters are learned in a single layer. Each filter represents a unique feature of the input data. Learning multiple features in different layers allows for the hierarchical feature construction of the input data \citep{zeiler2014visualizing}. 

The property of learning feature hierarchies makes CNNs suitable for classification of complex image data. CNNs have been very popular for computer vision and image processing applications \citep{lawrence1997face,krizhevsky2012imagenet}. Other applications in  astronomy where CNNs have been successfully used include galaxy morphology \citep{dieleman2015rotation}, star-galaxy classification \citep{kim2016star}, photometric redshift estimation \citep{hoyle2016measuring}, classification of optical transients \citep{cabrera2017deep} and classification of variable star light curves \citep{2017arXiv170906257M}, to name a few. The success rates of these applications have motivated us to use deep convolutional neural networks (DCNN) for detecting bars in galaxies. 

\section{Network Model}\label{sec:Network}
Neural networks are  designed according to the problem at hand. There are many hyper-parameters such as number of layers, size of kernels, weight initialization and many more that need to be considered for the design of a CNN. Even though no strict guidelines exist for the design, a simple model with two or three layers and minimal number of feature maps is designed first and then iteratively tweaked for better performance. This is a general convention when designing a network for a new problem. 

Initially, we explored different model architectures for this study, starting with two layers of convolutions and tried to tune the different hyper-parameters to reach an acceptable value of training loss and validation accuracy. The simple models failed to give reasonable accuracies (>90\%) during training. Afterwards, we increased the number of convolutional layers in a stacked fashion and also the number of feature maps within them. These additional modifications also did not prove to be better in classification accuracies. Further on, we chose to use a standard available network architecture for this study.

\begin{figure*} 
\centering
\includegraphics[scale=.5]{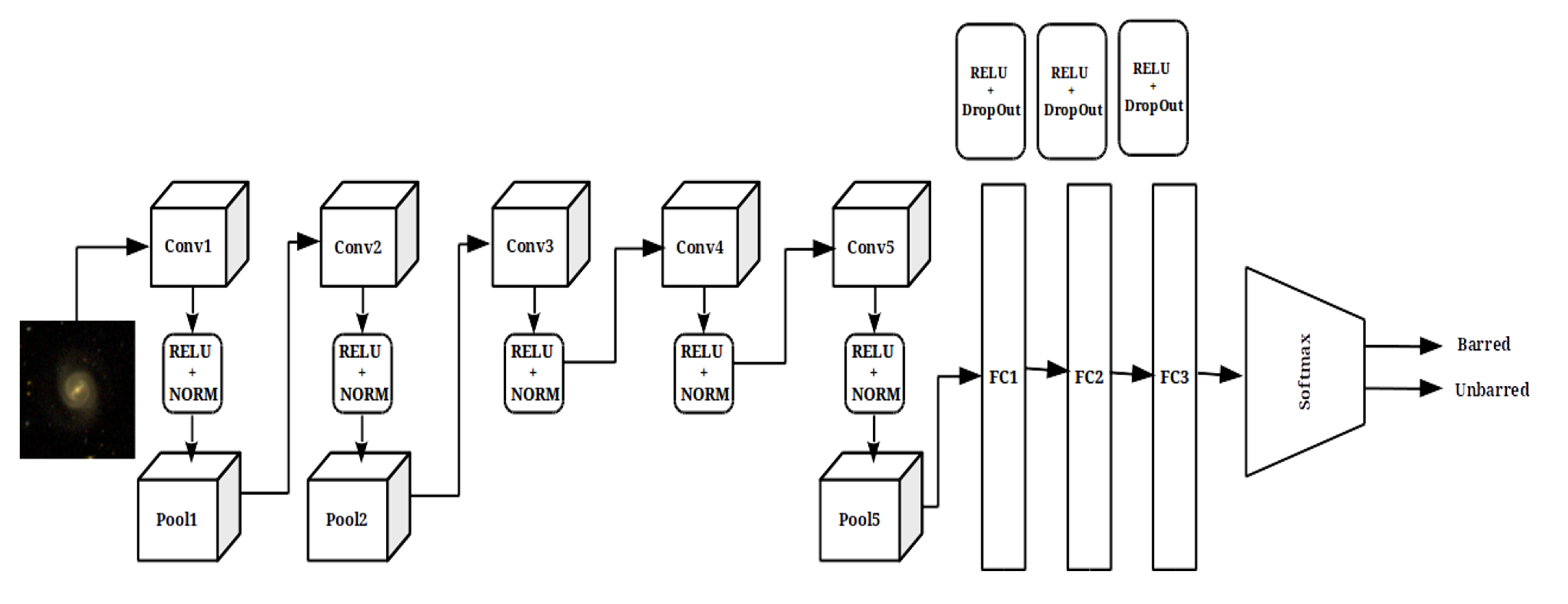}
\caption{Convolutional neural network used for this study. The network takes an RGB image as input and does the forward pass through a series of convolutions and finally gives the class probability scores as output.}
\label{fig:model}
\end{figure*}

We used the Alexnet convolutional neural network model which has been highly successful in different image recognition tasks \citep{krizhevsky2012imagenet}. It is one of the popular networks for complex image recognition tasks which performs with minimal loss and has comparatively less computational complexity. Alexnet was originally trained for the Imagenet challenge \citep{2014arXiv1409.0575R} which contained more than a million examples and 1000 classes. It is possible to use the pre-trained model for problems which have comparatively smaller training sets. But for our study, we opted not to use a pretrained model and trained a model from scratch. Figure~\ref{fig:model} shows the architecture of the DCNN model we adopted for this study.

\begin{figure}
\centering
\includegraphics[scale=.35]{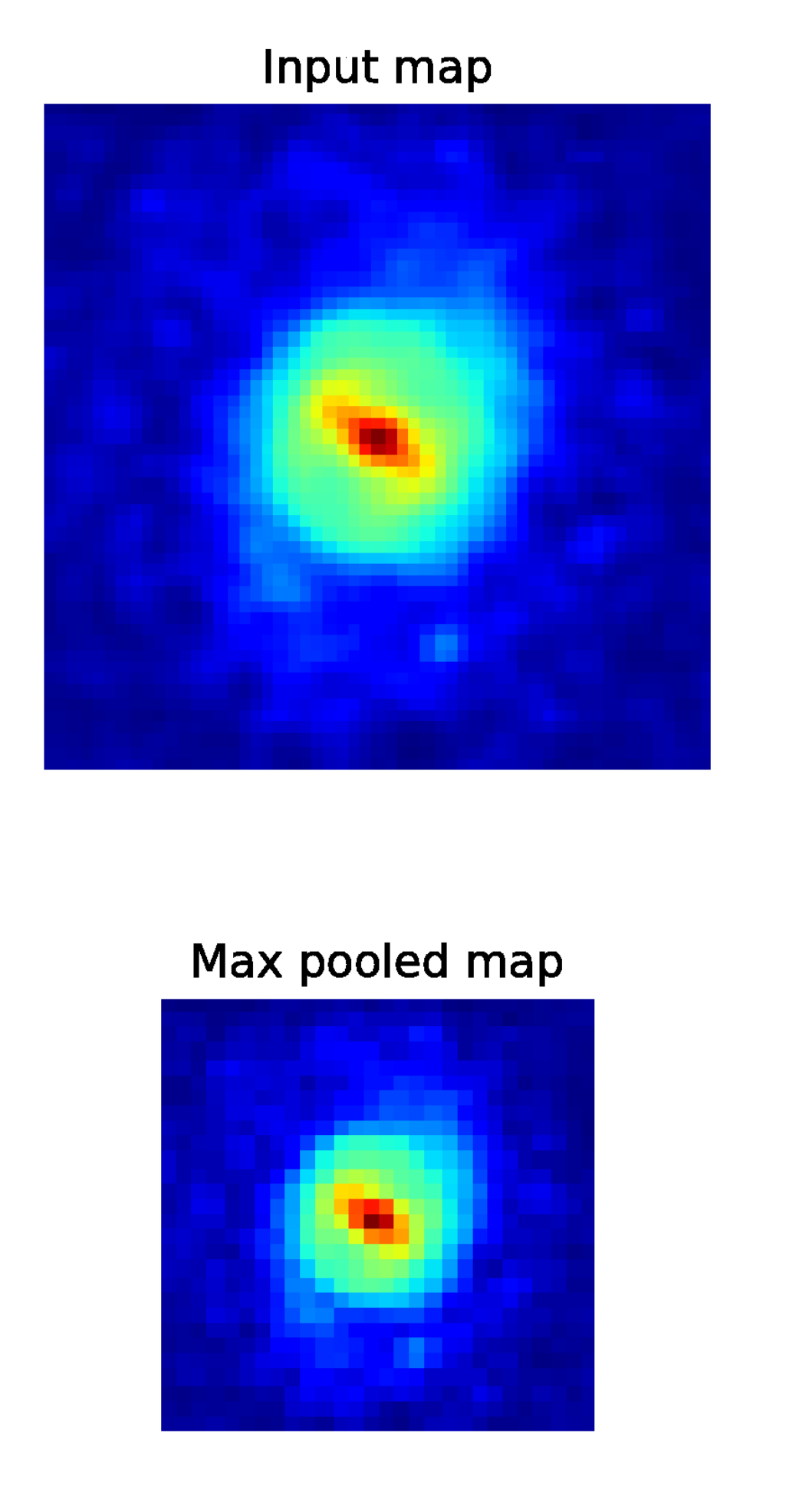}
\caption{Illustration of max-pooling operation is shown. The upper panel is an input map and the lower panel the max-pooled output which is the downsampled version. This operation reduces the computational complexity of the training. }
\label{pooling}
\end{figure}

The network has a total of 12 layers with five convolutional layers. The input to CNN is an image and the output from each CNN layer is called a feature map. The first two layers of convolutions are followed by a max-pooling layer which downsamples the output of the previous layer to half its original size. The first layer of convolution has 96 kernels of size $11 \times 11 \times 3$ and the output is fed to the second layer of convolution with 256 kernels each of size $5 \times 5 \times 48$ \citep{krizhevsky2012imagenet}. The max-pooling layers reduce the computational complexity and also help with learning rotational invariant features \citep{boureau2010theoretical}. The third, fourth and fifth convolutional layers learn more complex features in a serial fashion whose final output is fed into another max-pooling layer. Max-pooling is a subsampling method which downsamples an input map into half its size by finding the maximum value by sliding a $2 \times 2$ window over the input map. This is shown in Figure \ref{pooling}.

All the convolutional and pooling layers are followed by a Rectified Linear Unit activation \citep[ReLU;][]{nair2010rectified} and a batch normalisation (NORM) layer. The output of the final pooling layer is then fed into a series of fully connected layers which have 50 per cent dropout factor \citep{srivastava2014dropout}. Each of the fully connected layers also has a ReLU activation in each stage. The final layer is a softmax layer \citep{gold1996softmax} which computes the probability scores for the two classes. During training of the network the error in \ref{error} is minimized and the training loss is calculated from the cross entropy loss \citep{2006Sci...313..504H} which is a negative log likelihood and is given as \citep{jia2014caffe}
\begin{equation} \label{loss-eqn}
L(w) = - \frac{1}{N} \sum_{n=1}^{N} \left[y_{n} \log  \hat{y_{n}} + (1 - y_{n}) \log (1 -\hat{y_{n}}) \right]
\end{equation}
where $N$ is the number of training samples, $\mathit{w}$ is the weight vector, $\hat{y_{n}}$ the expected output and $y_{n}$ is output during a forward pass. The backpropagtion algorithm tries to minimize this loss function $L(w)$ during training.

The accuracy during the testing phase of each epoch is given as \citep{jia2014caffe}
\begin{equation}
\rm Accuracy = \frac{1}{N} \sum_{n}^{N} \delta \left\lbrace \hat{l_{n}} = l_{n} \right\rbrace ,   \delta \left\lbrace condition \right\rbrace 
\begin{cases}
1 & \text{if condition} \\
0 & \text{otherwise}
\end{cases}
\end{equation}
$N$ being the number of test samples, $\rm \hat{l_{n}}$ is the predicted class label for the $n^{th}$ sample and $\rm l_{n}$ is the true label. 

\section{Data Augmentation}\label{sec:Augm} 
Deep learning methods, in general, need a large number of training examples to learn and generate high accuracy results. The required number of training examples per class is of the
order of 10,000 or more. The rule is that the greater the number of examples the better will be the results. In cases where there are not enough examples, data augmentation through rotation and flipping of images helps generate enough examples and also solves bias issues \citep{krizhevsky2012imagenet}. This method is also called label preserving oversampling.

In this work,  we have only a few thousands of galaxy images from SDSS for training, which leads to high risk of over-fitting, i.e. the network models the training data too well and cannot generalise the model to unseen data. There are standard ways to evade over-fitting, and we used data augmentation by rotating the images of the galaxy. Since the rotation of an image does not affect the presence or absence of the bar feature, we rotate each image by one degree 359 times so that our augmented set becomes the final training sample which is 360 times larger than the original training set. Performing data augmentation by generating rotated versions of the same sample also has the following advantage. Convolutional neural networks have limited rotational invariance in the features they extract from the images \citep{krizhevsky2012imagenet}. Training the network with different rotated versions of the samples will help the network to extract fully rotational invariant features. Thus data augmentation helps us to increase vastly the training sample and also to make the network more rotationally invariant.

\section{Training and Testing}\label{sec:Train}
The labeled sample that we use in this study consists of 9346 galaxies of which 3864 are barred, and 5482 are unbarred. All the images were downloaded from SDSS DR13 with a cutout size of 300$\times$300 pixels in JPEG format.  The sample galaxies are divided into training and validation sets after random shuffle with a split ratio of 60 per cent for training and 40 per cent for validation.

\begin{figure}
\centering
\includegraphics[scale=.55]{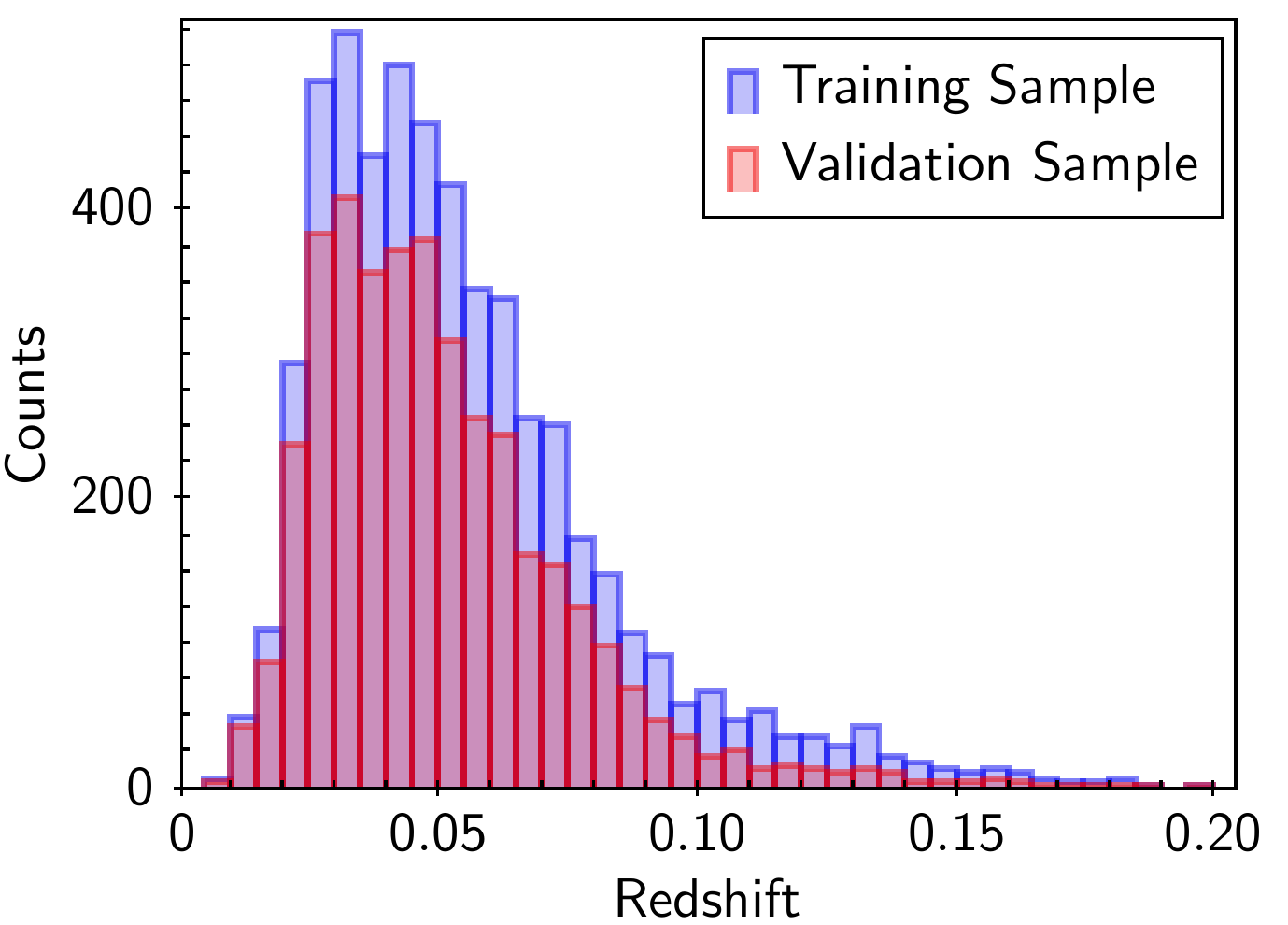}
\caption{The redshift distribution of training and validation samples is shown.} 
\label{fig:zdist_train}
\end{figure}

The split for the training sample was slightly adjusted to overcome the problem of class imbalance caused by the difference in the numbers of training samples for the barred and unbarred galaxies. 
For that, we have moved 10 per cent of the unbarred galaxies from the training set to the validation set and 10 per cent from the validation set to training set of barred galaxies. 
Thus there are 2704 barred and 2741 unbarred galaxies for the training set and 1157 barred and 2741 unbarred galaxies in the validation set. Figure~\ref{fig:zdist_train} shows the redshift distribution of the training and validation sample. 

\begin{figure}
\centering
\includegraphics[scale=.45]{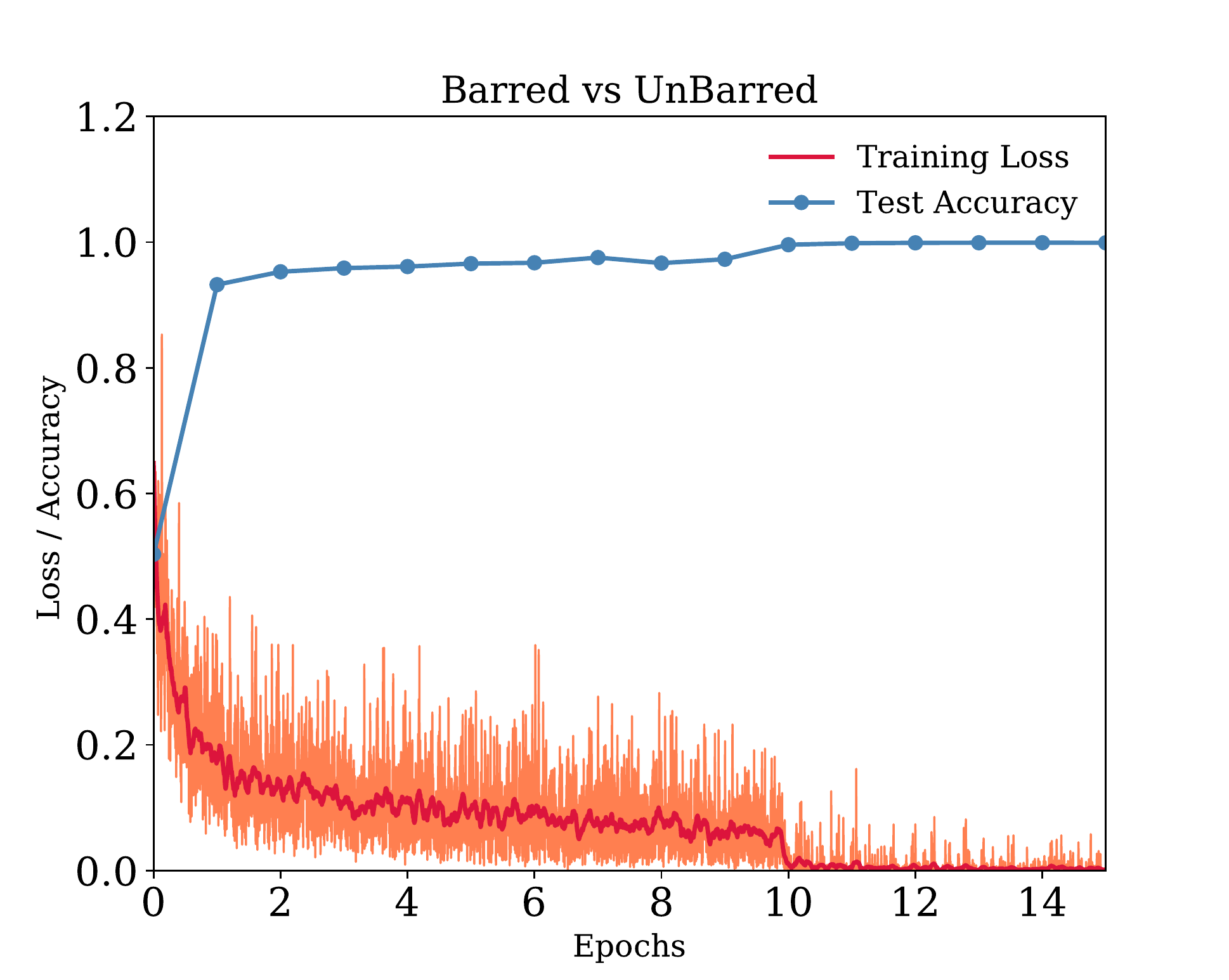}
\caption{The training loss and test accuracy of the model during training epochs. The red line shows the smoothed training loss over the training period. Each epoch of the model was tested against a test set whose accuracy is represented by the dotted blue line. It can be seen that the training loss converged to small values (<0.01) after 10 epochs.} 
\label{fig:curve}
\end{figure}

Depending on the complexity of the network architecture, convolutional networks have a significant number of learnable parameters, for example a million in the case of the network model used here \citep{krizhevsky2012imagenet}. This is the main reason why a convolutional neural network requires a substantial amount of data for training. As mentioned in the above Section~\ref{sec:occ}, we have rotated each galaxy image to generate additional training examples for the network. After each rotation, we cropped the image to a size of $256 \times 256$ pixels. Thus we have generated 1,960,200 training examples via rotation. Out of the augmented training examples, 80 per cent were taken to train the network, and 20 per cent for testing the accuracy during each epoch. For implementing the DCNN model, we chose the Caffe framework \citep{jia2014caffe}. All the training images in JPEG format were converted to Lightning Memory-Mapped Database (LMDB) format because it allows the data to be loaded into memory in a compressed format and also makes access much faster. The training was done on a server with an Intel(R) Xeon(R) CPU, 260 GB memory and four TITAN-Black GPUs with 12GB RAM each. We initially set the network to train for 50 epochs which in theory would take $\sim$ 31 hours to complete. A stochastic gradient algorithm with a base learning rate of 0.01 and a decay rate of 30 per cent was used for the training. The training loss and test accuracy tended to saturate after 10 epochs. It can be seen in Figure \ref{fig:curve} that during the initial epochs of training the loss was highly unstable because this is the phase where the network is still learning to extract the proper features. After this stage the loss saturates and training further may over-fit the model. With early stopping \citep{prechelt1998early} we were able to complete the training at around 16 epochs which took roughly 10 hours. In Figure~\ref{fig:curve}, the orange line which fluctuates rapidly shows the loss for each iteration. The very noisy appearance can be explained as follows. With a large number of training samples and a batch size of 100 examples used for forward and backward propagation, the number of iterations is also large. Therefore the loss is calculated for each iteration. An epoch is completed when all the training samples are forward propagated and weights of the network are updated. Thus the curve showing the loss for each iteration looks noisy in the learning curve plot. The red line in Figure \ref{fig:curve} is a smoothed approximation of the orange line.

\section{Network Analysis}\label{sec:occ}
In this section we analyze the network in terms of its working. Neural networks are currently limited by their tractability. Unlike other machine learning algorithms where extracted features are the input to the algorithm, CNNs take in raw data and generate features at each layer. For astronomical problems, one of the important considerations is to check if the network is learning features of the input data that have physical relevance. In the case of astronomical images, the cut-out may contain artefacts and even other unrelated sources like stars, galaxies etc. It is important to verify that the neural network is not learning features from such neighbouring  objects. In this study since the central bar feature is important, we performed a simple test to examine the ability of the network to identify genuine bars. This is explained below.

\subsection{Occlusion Test}
An obvious way to verify if the CNN is learning features from the bar in the galaxy is to perform an occlusion test. The general idea of this method is to check the prediction output by occluding the central bar in the galaxy images. Ideally, the CNN should classify a barred galaxy as unbarred when the central bar is masked. This will confirm whether the network is actually ``looking'' at the central bar.  

\begin{figure}
\includegraphics[width=0.45\textwidth]{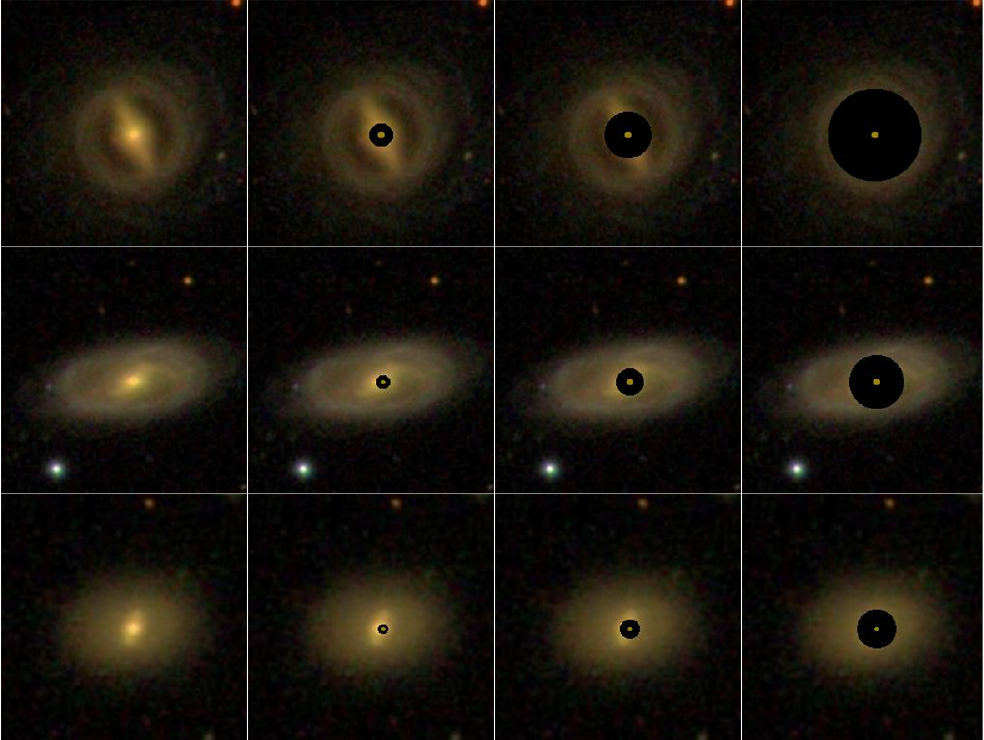}
\caption{Sample images showing the occlusion test for three galaxies with bars. The percentage of occlusion changes from 0 to 25, 50 \& 100 per cent respectively from left to right. We have seen that, when we mask the bar, the level of confidence in prediction by the network also decreases and the classification changes to unbarred when the level of masking is 50 per cent.} \label{fig:occlusion}
\end{figure}

The occlusion test is done in the following manner. A galaxy image cut out is chosen which has a prominent bar feature. A circular mask is placed over the image cutout such that it blocks the bar feature by 25, 50 and 100 per cent. We have chosen the physical distance from the center of galaxy so as to block the bar feature. In principle, when the bar is completely masked the prediction must change from barred to unbarred. The masking at different levels of occlusions is illustrated in Figure~\ref{fig:occlusion}.

We used a circular mask to occlude the galaxy bar because a rectangular mask may get detected as a bar from the initial layers of the network which have learned basic shapes to recognize a bar. Circular masks will not have this issue. In the occlusion test, we have found that the prediction changed from barred to unbarred when the occlusion level is at 50 per cent. This is also true when the central bar is completely masked. This shown in Table \ref{tab:ocul}. This proves that the CNN is learning features from the central bar of the galaxy.  

\begin{table}
\centering
\begin{tabular}{|c|c|}
\hline
\textbf{Level of Occlusion} & \textbf{Prediction} \\ \hline
25\%                        & Barred              \\ \hline
50\%                        & Unbarred            \\ \hline
100\%                       & Unbarred            \\ \hline
\end{tabular}
\caption{Table showing the prediction changing over different levels of occlusions presented to a barred galaxy.}
\label{tab:ocul}

\end{table}
  
\section{Results and Discussion}\label{sec:res}

Once the network model is trained, its performance has to be evaluated. The different measures of performance show the level of accuracy and biases present in the model. We tested our model with a validation set which the classifier has not seen during training. The different measures that we used to evaluate the validation results are as follows. 

Firstly, we calculate the precision and recall scores to find the accuracy of the model. Precision (P) and Recall (R) are defined as

\begin{equation*}
\hspace*{1.3cm}
\mathit{P} = \frac{\mathit{TP}}{\mathit{TP+FP}}, \hspace*{0.2cm} \mathit{R} = \frac{\mathit{TP}}{\mathit{TP+FN}}
\end{equation*}
where TP stands for true positive, TN for true negative, FP is false positive, and FN represents false negatives. In our case, TP is the number of barred galaxies correctly identified by the classifier, and TN is the number of unbarred galaxies correctly classified. Barred galaxies which are incorrectly identified as unbarred are false negatives and actual unbarred galaxies classified as barred galaxies by the network are the false positives. The recall, also known as the sensitivity of the classifier, can be used to check if the network is over-fitting. The precision and recall for the validation set are given in  Table~\ref{tab:conf}. We see that both have high values for barred and unbarred galaxies. These values are usually sensitive to the distribution of classes in the validation sample. Therefore a measure which is insensitive to the distribution should be used.

\begin{table}
\centering
\begin{tabular}{|c|c|c|c|}
\hline
 & Precision & Recall & Total No. of samples\\ \hline
Barred & 86.41 & 95.07 & 1157 \\ \hline
Unbarred  & 97.83 & 93.69  & 2741 \\ \hline
Overall & 94.1 & 94.1 & 3898  \\ \hline
\end{tabular}
\caption{The precision and recall for the validation set. The trained model is tested against a validation sample which has never been seen during the training process. It can be seen that the recall for both the barred and unbarred validation samples are very high. The slightly low value of individual precision for the barred galaxies is mainly due to the relatively smaller training samples compared to the unbarred samples.}
\label{tab:conf}
\end{table}

\begin{figure}
\centering
\includegraphics[scale =.5]{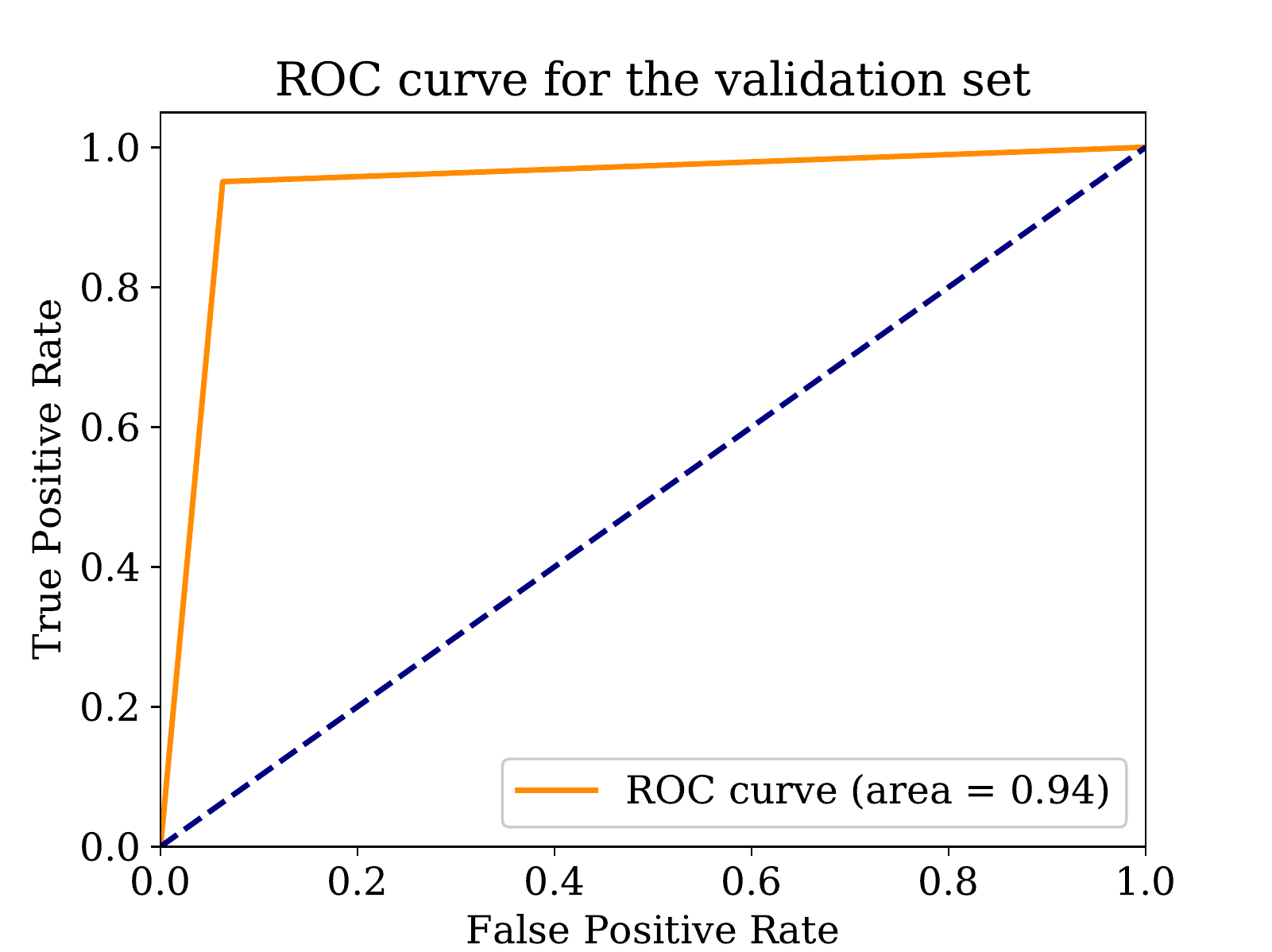}
\caption{Shown is the receiver operating characteristic (ROC) curve of predictions on the validation set. The dashed line shows the results for random guess. Ideally the curve should cover the whole region which just slightly larger than the current plot.}\label{fig:roc}
\end{figure}

Receiver Operating Characteristic (ROC) is one of the most commonly used methods for evaluating the performance of a binary classifier. The ROC curve shows the true positive rate against the false positive rate for the validation sample. One of the important properties of the ROC curve is that it is insensitive to class distribution. In our validation set the sample distribution of the two classes were not exactly equal. Therefore the ROC curve is well suited to evaluate our trained model. The area under the curve (AUC) represents the accuracy of the classifier from the ROC curve. Figure~\ref{fig:roc} represents the ROC curve for the predictions on the validation set. In our case the AUC is 0.94 which is very close to the ideal value of 1 in the normalized scale. 

\begin{figure}
\centering
\includegraphics[scale =.5]{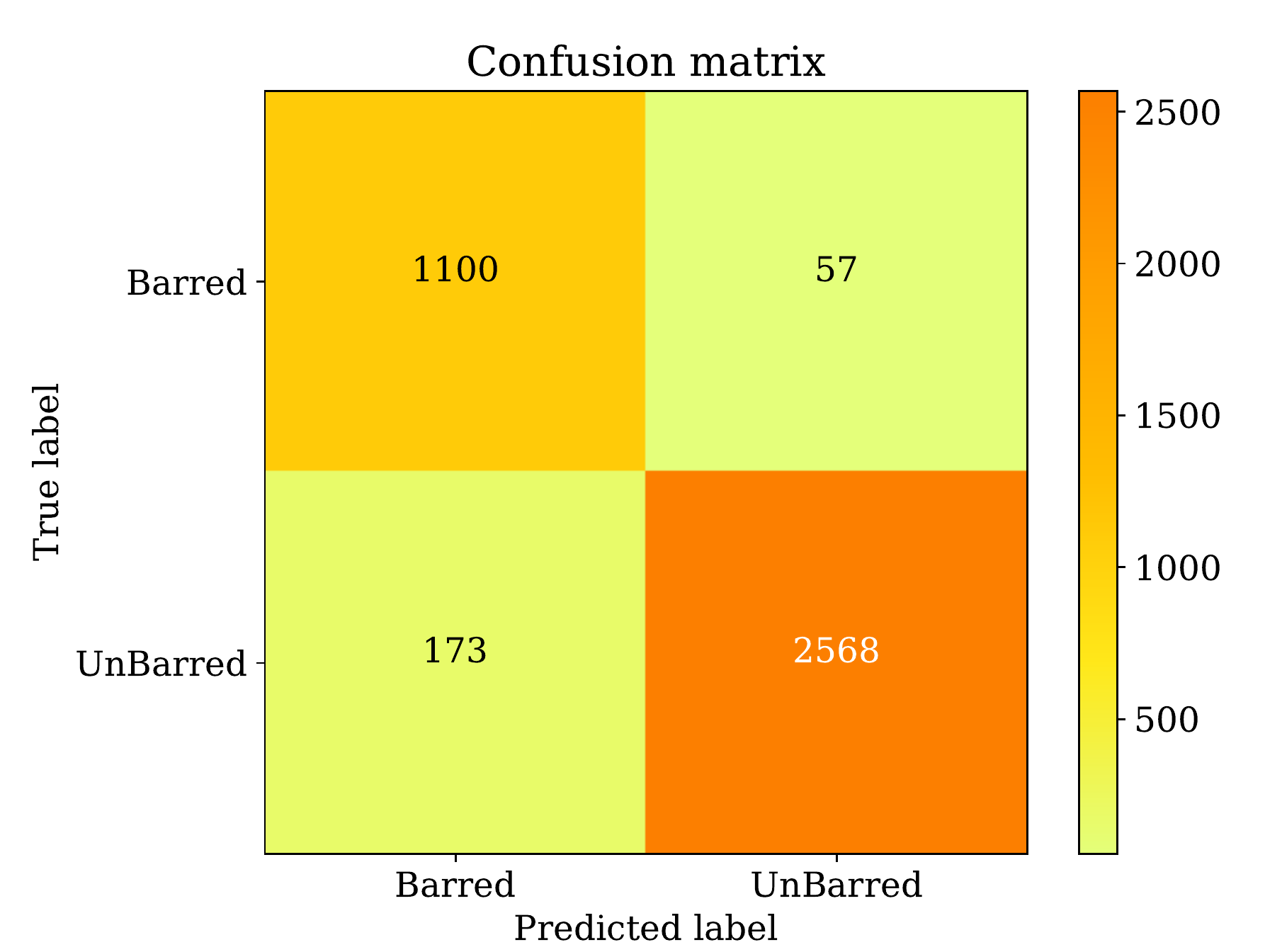}
\caption{Confusion matrix of predictions showing how many of the positive classes were correctly and incorrectly predicted by the classifier.}
\label{fig:conf}
\end{figure}

The somewhat lower value for individual precision of 86.41 per cent for the detection of barred galaxies is mainly because the number of barred galaxies in the training set was less than the number of unbarred galaxies. Broadly speaking this will have reduced the representativeness of the barred training examples, affecting the precision of the model. But in the case of unbarred galaxies, both precision and recall have high values of  97.83 per cent and 93.69 per cent respectively. This shows that the model was fairly balanced in the case of learning unbarred galaxies. This can be further explained with numbers from the confusion matrix, in Figure~\ref{fig:conf} which shows the correct and incorrect predictions for each class in the validation set. 

\begin{figure}
\centering
\includegraphics[scale =.8]{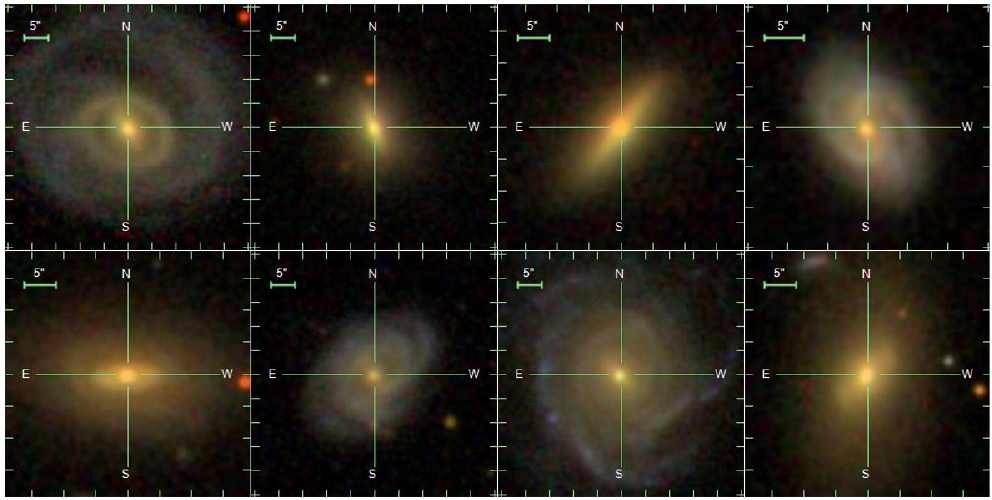}
\caption{Sample galaxy images which are incorrectly predicted by our network model. The upper panel shows the unbarred galaxy samples in our validation set predicted as barred by the network. Barred galaxies identified as unbarred are shown in the lower panel. Both the samples look visually similar, which could have confused the network.}
\label{fig:failed}
\end{figure}

It can be seen from Figure~\ref{fig:conf} that 173 unbarred galaxies are classified as barred galaxies. We found that most such galaxies had close resemblance to barred galaxies which had short-faint bars. In most cases, the central bulge looked similar to the short-faint bars seen at the center of a barred galaxy. A few such examples where the network may have got confused and classified unbarred galaxies as barred are shown in the upper panel of Figure \ref{fig:failed}. 

In the case of barred galaxies being classified as unbarred, most of the failures were mainly because the central region of the galaxy was too bright. This makes it difficult for the network to clearly identify the bar feature from the bright area. In some cases the bars were either too faint or short to be detected. A few such misclassified barred galaxies are also shown in the lower panel of Figure~\ref{fig:failed}, where we can see that the failed samples in both cases are visually similar. This means that they are close to each other near the class decision boundary in the feature space of the model.

One way to reduce the effect of the faint bar resembling feature in unbarred galaxies is to perform appropriate preprocessing of the images. We experimented with general methods like contrast adjustments and sigma-clipping  in the initial stages of our work and found that the results were not satisfactory. This is mainly because such steps were not sensitive enough to highlight the bar feature. 

One of the features of the galaxy images that may have confused the network is a bright central bulge. To see if the effect of such a bright central bulge in the images could be reduced, we tested two methods. In the first method the mean image of all the training examples was obtained and subtracted from all the galaxy images which were used for training and testing. In the second method, for each galaxy, we obtained the mean pixel value over a small central region of the galaxy and subtracted this mean value from the galaxy image. We then trained the network separately using  these two samples. For both the cases, the training accuracy was less than 80 per cent, showing that neither was an effective strategy in our case.

Background stars, galaxies or artefacts could lead to false identification as a barred galaxy. To try to reduce the effect of such background sources, we trained the network by rescaling the image so that the entire galaxy image is confined to 256x256 pixels. But this did not help to improve the accuracy of the classifier. This may have been ineffective because the network focuses on the centre of the galaxy as explained in Section \ref{sec:occ}. Another reason could be that we have removed images with numerous artefacts from our sample, so that the technique is inefficient for this data set but may be useful when we deal with a larger amount of data. Therefore we have used the raw images directly by incorporating the selection criteria mentioned in Section~\ref{sec:data}.  

\section{Catalogue and Web Interface}\label{sec:cat}
We have generated a catalogue of barred galaxies using the predictions from our trained network. The prediction data is taken from SDSS DR13 which has 169,616 galaxies which satisfy all our selection criteria mentioned in Section \ref{sec:data}. We  cross-matched this data with the \citet{2015MNRAS.446.3943M} catalogue which contains the 2-D decomposition of $\sim 7 \times 10^{5}$ spectroscopically selected galaxies from SDSS DR7. The catalogue has various structural and morphological parameters such as surface brightness of bulge and disk, bulge effective radius, disc scale length etc. But the catalogue does not have any bar information incorporated. So we chose the galaxies from \citet{2015MNRAS.446.3943M} catalogue.  The cross-matching resulted in 111,838 galaxies in common. The catalogue constructed in this work is based on this cross-matched samples. 

\begin{figure}
\centering
\includegraphics[scale =.8]{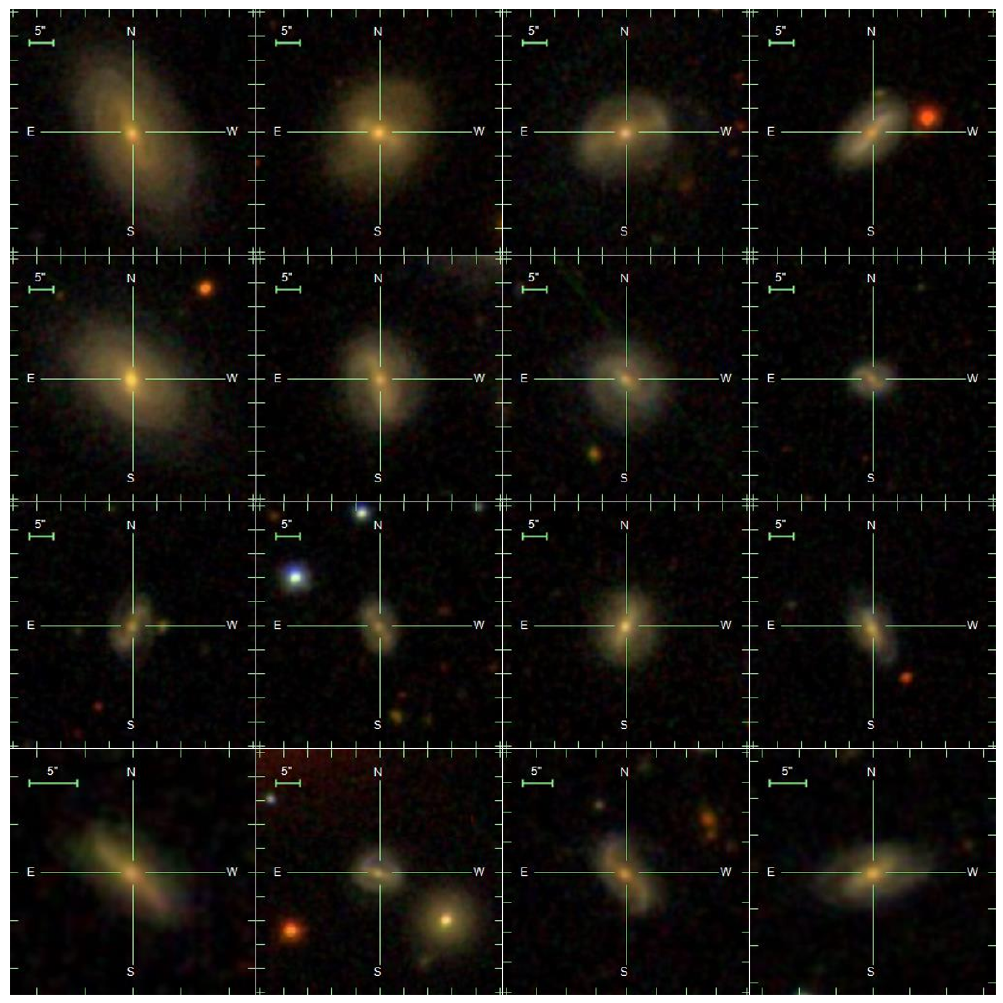}
\caption{A sample of galaxies predicted as barred by our network.  }
\label{fig:predbar}
\end{figure}

\begin{figure}
\centering
\includegraphics[scale =.5]{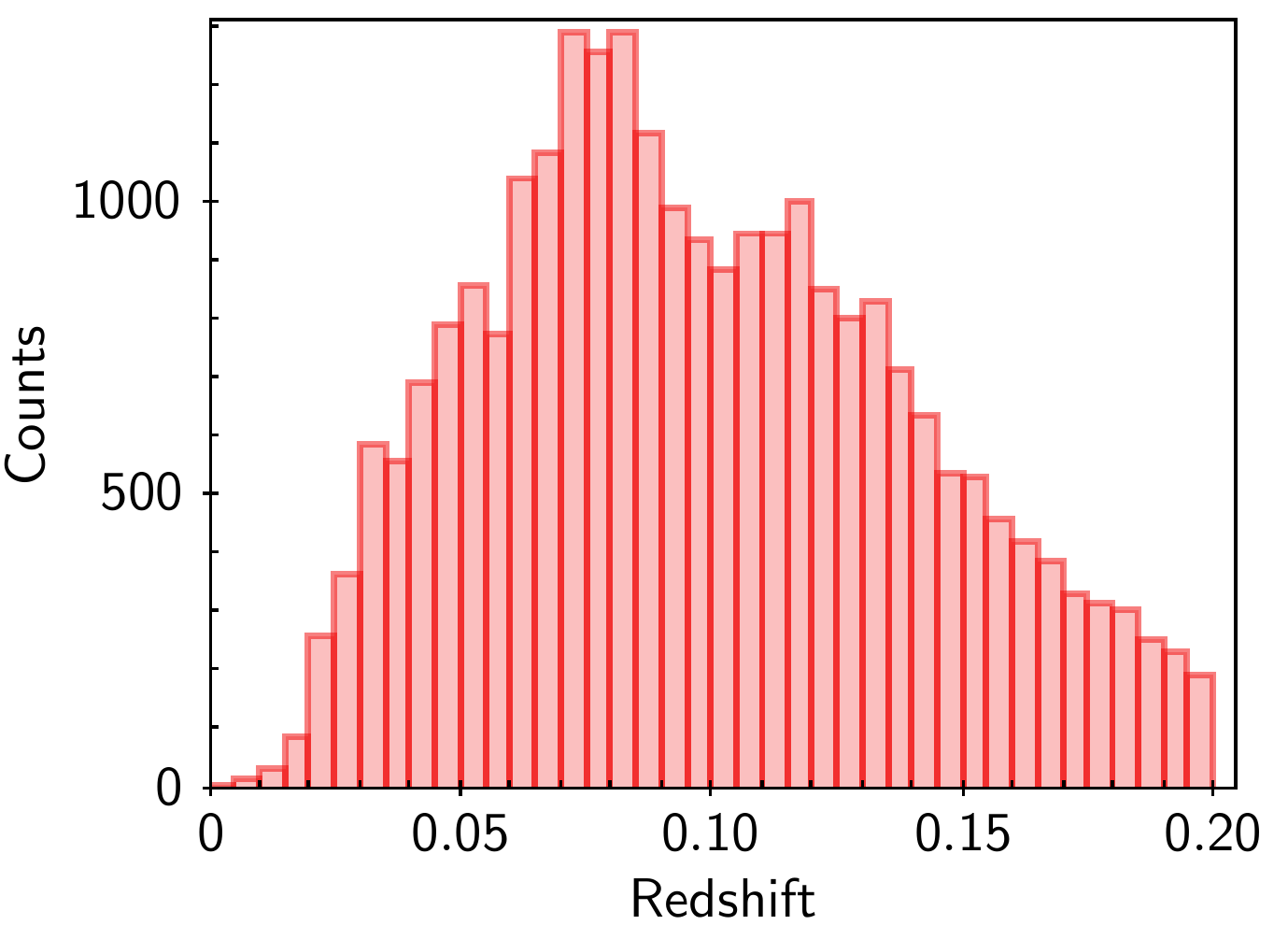}
\caption{The redshift distribution of galaxies in our catalogue predicted as barred by the trained CNN classifier.} \label{fig:zdist}
\end{figure}

The network predicted 25,781 galaxies as barred in our catalogue. The distribution of redshift of the galaxies predicted as barred in our catalogue is shown in Figure~\ref{fig:zdist}. Out of the 111838 galaxies in the prediction sample, 53154 galaxies have Galaxy Zoo classifications, and the rest do not have classification available in Galaxy Zoo data release 2. In Figure~\ref{fig:predbar}, we have shown some of the galaxies predicted as barred by our network which do not have galaxy zoo classification.  There are 4803 barred galaxies according to Galaxy zoo classification, and our network successfully identified 2695 galaxies as barred. Most of the remaining galaxy zoo barred galaxies have faint and diffuse bar-like structure at their centre. While training the network, we had removed all galaxy images with faint, diffuse bar-like feature and this may be the reason why the network has missed these galaxies. Table~\ref{tab:cat} shows sample rows from our catalogue. The entire catalogue of 111,838 galaxies with predictions about the presence of bar is available as an online material. 

\begin{table*}
\centering
\begin{tabular}{|c|c|c|c|c|}
\hline
SDSS\_Objid & ra & dec & Class & Probability \\ 
& (deg) & (deg) & & (per cent) \\ \hline
1237656496187441273 & 0.0058 & 15.50979 & Unbarred & 99.9998 \\
1237656496724312079 & 0.0088 & 15.88179 & Unbarred & 100.0 \\
1237657191978959126 & 0.0198 & 0.78170 & Barred & 87.3507 \\
1237652944249880792 & 1.9431 & 15.73540 & Unbarred & 55.4469 \\ 
1237649920574488695 & 10.6244 & 15.39671 & Barred & 99.9946 \\
1237653653450915964 & 12.6296 & 16.08139 & Unbarred & 99.9995 \\
1237649921112277085 & 12.7767 & 15.88129 & Barred & 99.9916 \\
1237672795035140280 & 3.0638 & -8.98779 & Unbarred & 99.9868 \\
1237652900229742736 & 42.7542 & -8.19610 & Unbarred & 100.0 \\
1237651273490694741 & 114.2101 & 39.60299 & Barred & 99.9998 \\ \hline
\end{tabular}
\caption{Table showing the sample rows from our catalogue based on SDSS DR13. Column 1 is the SDSS photometric object ID, ra - right ascension in decimal degrees (J2000), dec - declination in decimal degrees (J2000), Class - most probable class of the object and Probability - the prediction probability for the most probable class. The full table is available online.}
\label{tab:cat}

\end{table*}

We have created a web interface for our trained network for public use. The interface can be accessed via the URL: http://ddi.iucaa.in/barClassifier. The web interface is easy to use and allows the user to upload multiple images. Once the user uploads the images and submits a request, the images are classified on the server and a table is output to the user containing a preview thumbnail, the name of the image file, a flag indicating whether the galaxy is barred or not and the probability associated with the same. The user can download the result table in ASCII format.

The web application mentioned also contains an FAQ to guide the user about the service and contains links to a) the trained neural network files and b) the classification code and c) the catalog of 111,838 galaxy classifications described in this paper. We are planning to incorporate different additional features to this website in the future. 

\section{Conclusions}\label{sec:concl}
We have demonstrated the usefulness of a deep convolutional neural network for identifying bars in galaxies from SDSS  images with a top precision of 94 per cent. The main advantage of using a convolutional neural network is the automatic extraction of features from the raw images. With a base sample size of a few thousand, we have made use of the rotated images of example galaxies to generate the large amount of data required for training the model. By this bootstrapping of the sample, we were able to train the model to high accuracy. The trained classifier is used to construct a catalogue of barred galaxies from SDSS DR13. There are 25781 barred galaxies in this catalogue which is available online through the web interface mentioned in Section~\ref{sec:cat}. 

In this study we have not attempted to classify the different types of bars in galaxies. This was more of an attempt to show the utility of deep learning for an interesting and challenging problem with galaxy morphology. Even though the samples used in this study mainly consisted of a mixture of strong, weak and short barred galaxies, our network was able to identify specific types like nuclear and peanut shaped bars present in the validation set. Another advantage of the convolutional neural network is that this same model can be used as a feature extractor for studying the different types of bars. Also the model can be retrained to detect those specific types without the need of having a large training set. This kind of retraining is often known as transfer learning \citep{bengio2012deep}.  

The lack of a good representative sample set which is large enough to train a deep convolutional neural network is a common problem faced in such studies. Generally, galaxy morphological features are identified by humans and so they suffer from subjective biases. Our training sample also suffers from the same bias, because the bars were visually identified ones from the galaxy zoo project and by \citet{2010ApJS..186..427N}. As a future extension to this study we intend to improve the model by using a better representative training sample and one which has very little human bias. This will ultimately help in improving the accuracy of the model by reducing the number of false positives.

Deep learning is gaining popularity in many areas of astronomy mainly because it allows to bypass the hurdle of feature engineering and provides exceptionally accurate results. We have demonstrated this utility in the present work by detecting bars in galaxies. Automating this process with convolutional neural networks is highly efficient in terms of computational cost because it directly uses images and requires only a fraction of a second for the classification.

\section*{Acknowledgements}
The authors express thanks to Aniruddha Kembhavi for a careful reading of the manuscript and comments. A.K.Aniyan would like thank the Square Kilometer Array South Africa (SKA SA) for funding the research project. This research has been conducted using resources provided by the Science and Technology Facilities Council (STFC) through the Newton Fund and the SKA Africa. 

Funding for the Sloan Digital Sky Survey IV has been provided by the Alfred P. Sloan Foundation, the U.S. Department of Energy Office of Science, and the Participating Institutions. SDSS-IV acknowledges support and resources from the Center for High-Performance Computing at the University of Utah. The SDSS web site is www.sdss.org. 

SDSS-IV is managed by the Astrophysical Research Consortium for the 
Participating Institutions of the SDSS Collaboration including the Brazilian Participation Group, the Carnegie Institution for Science, 
Carnegie Mellon University, the Chilean Participation Group, the French Participation Group, Harvard-Smithsonian Center for Astrophysics, Instituto de Astrof\'isica de Canarias, The Johns Hopkins University, Kavli Institute for the Physics and Mathematics of the Universe (IPMU) / University of Tokyo, Lawrence Berkeley National Laboratory, Leibniz Institut f\"ur Astrophysik Potsdam (AIP),  Max-Planck-Institut f\"ur Astronomie (MPIA Heidelberg), 
Max-Planck-Institut f\"ur Astrophysik (MPA Garching), 
Max-Planck-Institut f\"ur Extraterrestrische Physik (MPE), 
National Astronomical Observatories of China, New Mexico State University, New York University, University of Notre Dame, 
Observat\'ario Nacional / MCTI, The Ohio State University, 
Pennsylvania State University, Shanghai Astronomical Observatory, 
United Kingdom Participation Group, Universidad Nacional Aut\'onoma de M\'exico, University of Arizona, University of Colorado Boulder, University of Oxford, University of Portsmouth, University of Utah, University of Virginia, University of Washington, University of Wisconsin, Vanderbilt University, and Yale University.



\bibliographystyle{mnras}
\bibliography{ref} 



\label{lastpage}
\end{document}